\documentclass[aps,prl,twocolumn,superscriptaddress,showpacs]{revtex4}
\usepackage{amsmath,amssymb,bm}
\usepackage{graphicx,epsfig}

\newcommand{\be}{\begin{equation}}
\newcommand{\ee}{\end{equation}}
\newcommand{\nn}{\nonumber \\}
\newcommand{\ij}{\langle ij \rangle}

\newcommand{\ba}{\begin{eqnarray}}
\newcommand{\ea}{\end{eqnarray}}
\newcommand{\Abar}{A^{\dag}}

\newcommand{\abar}{a^{\dag}}

\newcommand{\Ubar}{U^{\dag}}

\newcommand{\bpm}{\begin{pmatrix}}
\newcommand{\epm}{\end{pmatrix}}
\newcommand{\bbar}{b^\dag}

\newcommand{\kbar}{\overline{k}}
\renewcommand{\v}[1]{\textbf{#1}}

\begin{document}

\title{Vector chiral states in low-dimensional quantum spin systems}
\author{Raoul Dillenschneider}
\affiliation{Department of Physics, BK21 Physics Research Division,
Sungkyunkwan University, Suwon 440-746, Korea}
\author{Junghoon Kim}
\affiliation{Department of Physics, BK21 Physics Research Division,
Sungkyunkwan University, Suwon 440-746, Korea}
\author{Jung Hoon Han}
\email[Electronic address:$~$]{hanjh@skku.edu}
\affiliation{Department of Physics, BK21 Physics Research Division,
Sungkyunkwan University, Suwon 440-746, Korea} \affiliation{CSCMR,
Seoul National University, Seoul 151-747, Korea}

\begin{abstract}
A class of exact spin ground states with nonzero averages of vector
spin chirality, $\langle \v S_i \times \v S_j \cdot \hat{z} \rangle$,
is presented. It is obtained by applying non-uniform O(2) rotations
of spin operators in the XY plane on the SU(2)-invariant
Affleck-Kennedy-Lieb-Tasaki (AKLT) states and their parent
Hamiltonians. Excitation energies of the new ground states are
studied with the use of single-mode approximation in one dimension
for $S=1$. The excitation gap remains robust. Construction of chiral
AKLT states is shown to be possible in higher dimensions. We also
present a general idea to produce vector chirality-condensed ground
states as non-uniform O(2) rotations of the non-chiral parent states.
Dzyaloshinskii-Moriya interaction is shown to imply non-zero spin chirality.
\end{abstract}

\maketitle

\textbf{Introduction}: Vector spin chirality, defined as the
projection onto the axis of rotation (here given by $\hat{z}$) of the
average of the outer product of two adjacent spins, $ \kappa_{ij} =
\langle \hat{\kappa}_{ij} \rangle$, $\hat{\kappa}_{ij}=\v S_i \times
\v S_j \cdot\hat{z}$, measures the sense of rotation of the magnetic
moments in a spiral magnet. Being even under time reversal and odd
under the inversion of $i$ and $j$ sites, this chirality plays an
important role in the recent study of spin-polarization coupling in
multiferroic materials where the local dipole moment shares the same
symmetry properties as $\kappa_{ij}$ \cite{theory-of-MF}. A linear
coupling between the two order parameters is a generic phenomenon in
spiral magnets.


In a quite different context, an interesting observation was made by
Hikihara \textit{et al}. for the $S=1$ spin chain with both nearest
($J_1$) and next-nearest $(J_2)$ neighbour
interactions \cite{hikihara}. For $J_2 /J_1$ larger than a critical
ratio, the ground state was shown to possess long-range order in the
chirality correlation function, $\langle \hat{\kappa}_{i,i+1}
\hat{\kappa}_{j,j+1} \rangle$, as $|i-j|\rightarrow \infty$. Such a
novel phase, in the context of multiferroicity, would result in a
strong coupling to the local dipole moment even in the absence of
magnetic order. A similar possibility of a non-magnetic, yet
chirality-ordered phase was explored in the Ginzburg-Landau treatment
of anisotropic spin models\cite{onoda}. In both instances, the key is
to reduce the symmetry of the Hamiltonian away from SU(2) and
introduce frustration to suppress magnetic ordering.

In this paper, we discuss a simple route to produce the vector chiral
ground state in low-dimensional spin systems.
%
%
First we look to models of Heisenberg spin exchange
together with the Dzyaloshinskii-Moriya (DM) interaction $
\sim \v S_i \times \v S_j \cdot \hat{z}$. In the
one-dimensional case, one can consider the following
Hamiltonian:

\be H[\{\theta_i \}] = J_1 \sum_{\ij} S^z_i S^z_j \!+\! J_2
\sum_{\ij} \left( e^{i\theta_{ij}} S^+_i S^-_j \!+\!
e^{-i\theta_{ij}} S^+_j S^-_i \right). \label{1D-with-DM}\ee
We choose $i=j+1$ and  $S_i^{\pm} = (S^x_i \pm i S^y_i
)/\sqrt{2}$, for a periodic lattice of length $N$. The
strength of the DM interaction for each $\ij$ bond is given
by $J_2 \sin \theta_{ij}$. One can implement a
site-dependent unitary rotation of the spins, $S^+_i
\rightarrow U_i S_i^+ \Ubar_i = S_i^+ e^{-i\theta_i}$, with
the angle $\theta_i$ chosen to meet the condition
$e^{i(\theta_i - \theta_j )}  =
e^{i\theta_{ij}}$\cite{aharony}. A simple way to choose
$\theta_i$ is to start with $\theta_1 = 0$, then choose all
successive angles as $\theta_2 = \theta_{21}$, $\theta_3
-\theta_2 = \theta_{32}$, etc. according to $\theta_i =
\sum_{2 \le j\le i}\theta_{j,j-1}$. Due to the periodic
structure we need to require $\theta_{N+1} $ ($\equiv
\theta_1 $), which equals the sum of all the bond angles
$\sum_{1\le 1 \le N} \theta_{i,i-1}$, be an integer
multiple of $2\pi$:

\be \sum_{1\le i \le N} \theta_{i,i-1} /2\pi = n
(=\mathrm{integer}).\label{integer-flux}\ee
Once this condition is met, it is always possible to ``gauge away"
the phase angles in the model given in Eq. (\ref{1D-with-DM}) to
reduce it to the XXZ Hamiltonian: $U[\{\theta_i \}]H[\{\theta_i \}]
\Ubar [\{\theta_i \}] =H_{\mathrm{XXZ}}$ where $U[\{\theta_i \}] =
\prod_i U_i $. The eigenstates of Eq. (\ref{1D-with-DM}),
$|\{\theta_{ij}\}\rangle$, have a one-to-one correspondence with
those of the XXZ model, denoted $|\mathrm{XXZ} \rangle$, and given
explicitly by $|\{\theta_{ij}\}\rangle =\Ubar [\{\theta_i
\}]|\mathrm{XXZ}\rangle$.

Symmetry consideration dictates that  $\langle  S^-_i S^+_j \rangle$
for the  eigenstates of $H_{\mathrm{XXZ}}$ be equal to $X_{ij}$,
where $X_{ij}$ is a real-valued number. It then follows that $\langle
\{\theta_{ij}\} |S^-_i S^+_j |\{\theta_{ij}\} \rangle =X_{ij}
e^{i(\theta_i -\theta_j ) }$ for the eigenstates $|\{\theta_{ij}\}
\rangle$. The imaginary part of this average is nothing but the spin
chirality, $\langle S_i^x S_j^y \!-\! S_i^y S_j^x \rangle$, given by
$X_{ij} \sin (\theta_i -\theta_j )$. This simple argument proves that
the DM interaction induces non-zero spin chirality in the quantum
eigenstates.

For $S=1/2$, through Jordan-Wigner transformation, the Hamiltonian
(\ref{1D-with-DM}) is mapped to a model of spinless fermions coupled
to the gauge flux $\theta_{ij}$. While a persistent current will
exist for general values of the flux $\phi=\sum_{i} \theta_{i,i-1}$,
the criteria given in Eq. (\ref{integer-flux}) corresponds to having
an integer multiple of the flux quantum threading the ring, for which
we would expect vanishing fermion current. Here, however, one must
note that the spin chirality maps onto $i\langle f^+_i f_j \!-\!
f^+_j f_i \rangle$, whereas the gauge-invariant definition of the
fermion current will be $i\langle e^{i\theta_{ij}} f^+_i f_j \!-\!
e^{-i\theta_{ij}}f^+_j f_i \rangle$. This latter quantity vanishes
when the flux is an integer multiple of $2\pi$ but the spin
chirality, given by $i\langle f^+_i f_j \!-\! f^+_j f_i \rangle$ in
the fermion language, remains nonzero even for the integer flux case.
In turn, $\langle e^{i\theta_{ij}}S^+_i S^-_j \!-\!
e^{-i\theta_{ij}}S^-_i S^+_j\rangle $ vanishes for
$|\{\theta_{ij}\}\rangle$ when an integer flux threads the ring. Our
proof remains valid for arbitrary spin $S$ (for which no
Jordan-Wigner transformation exists) and $J_2 /J_1$ ratio.

The whole class of Hamiltonians given by Eq.
(\ref{1D-with-DM}) obeys an identical set of energy spectra
regardless of the choice of bond angles $\{\theta_{ij}\}$,
as long as Eq. (\ref{integer-flux}) is obeyed. We have
checked this for a 4-site model with arbitrary
$\{\theta_{12},\theta_{23},\theta_{34}, \theta_{41} \}$,
under the constraint
$\theta_{12}+\theta_{23}+\theta_{34}+\theta_{41}=2\pi
\times \mathrm{integer}$, for both $S=1/2$ and $S=1$ cases.
The spin-spin correlation functions also behave in the
manner predicted by the gauge argument.


Reversing the argument, one can generate states of non-zero and
non-uniform chirality beginning with the XXZ Hamiltonian by
introducing a site-dependent phase angle $\theta_i$ and rotating each
spin accordingly:  $S^+_i \rightarrow \Ubar_i S_i^+ U_i = S_i^+
e^{i\theta_i}$. The XXZ Hamiltonian undergoing the unitary rotation
$\Ubar [\{\theta_i \}] H_{\mathrm{XXZ}} U[\{\theta_i \}]=H[\{\theta_i
\}]$ becomes Eq. (\ref{1D-with-DM}) with $\theta_{ij} = \theta_i -
\theta_j$. To obtain the uniform DM interaction one can use $\theta_i
= \theta \times i$ where $i$ is the local coordinate and require that
$\theta N$ ($N$=number of lattice sites) be an integer multiple of
$2\pi$. For the staggered DM interaction one can choose $\theta_i =
0$ and $\theta$ for even and odd sites, respectively. The net flux,
given by Eq. (\ref{integer-flux}), will be always zero for even $N$,
regardless of $\theta$. The eigenstates, obtained as unitary
rotations of those of the XXZ Hamiltonian, will have non-zero spin
chirality.
\\

\textbf{One-dimensional chiral AKLT state}: We have presented an
argument how a quantum state with non-zero spin chirality can be
generated. The same idea can be applied to the well-known
Affleck-Kennedy-Lieb-Tasaki (AKLT) ground states of spins for one
dimension\cite{AKLT}. The discussion is most conveniently carried out
in the Schwinger boson language where the spin operators are
represented by $S_i^+ = \abar_i b_i /\sqrt{2}$, $S_i^- = \bbar_i a_i
/\sqrt{2}$ and $S_i^z = (\abar_i a_i - \bbar_i b_i )/2$. Spin
rotation in the XY plane is implemented through $\abar_i \rightarrow
\abar_i e^{i\theta_i/2}$, and $\bbar_i \rightarrow \bbar_i
e^{-i\theta_i/2}$. Under this rotation, the AKLT ground state, which
is built up of a product of bond singlet operators $\Abar_{ij} =
\abar_i \bbar_j - \bbar_i \abar_j$, is replaced by

\be  |\{ \theta_i \} \rangle \!=\! \prod_{\ij} \Abar_{ij}
[\theta_{ij}] |0 \rangle, \label{AKLT-wf}\ee
where $\Abar_{ij} [\theta_{ij}] = e^{i\theta_{ij}/2} \abar_i \bbar_j
- e^{-i\theta_{ij}/2}\bbar_i \abar_j $, and $\theta_{ij} = \theta_i
\!-\! \theta_j$. The quantization rule shown in Eq.
(\ref{integer-flux}) is satisfied. The AKLT ground state can be
written in the matrix product form\cite{zittartz},
$|\mathrm{AKLT}\rangle = \mathrm{Tr}\left( \prod_i g_i \right)$ with
a $2\times 2$ matrix $g_i$, and we can write down a similar matrix
product ground state for non-zero chirality, $|\{\theta_i \} \rangle
= \mathrm{Tr}\left( \prod_i g_i [\theta_i] \right)$, using

\be g_i [\theta_i ] = \left( \begin{array}{cc} \abar_i \bbar_i &
- e^{i\theta_i} (\abar_i )^2 \\
e^{-i \theta_i } ( \bbar_i )^2  & - \abar_i \bbar_i
\end{array}
\right)  .\ee
As will be shown shortly, the states given in Eq. (\ref{AKLT-wf})
show non-zero spin chirality, and may be christened the ``chiral AKLT
states". The AKLT Hamiltonian undergoes the unitary rotation
accordingly. Expressed in the Schwinger boson language,
$H[\{\theta_{ij} \}] = \sum_{\ij} H_{ij} [\theta_{ij}] $, the
pair-wise Hamiltonian $H_{ij}[\theta_{ij}] $ is given as

\ba && H_{ij} [\theta_{ij} ] = \frac{1}{24} (6 \!-\!
A^\dagger_{ij}[\theta_{ij}] A_{ij}[\theta_{ij}] ) (4 \!-\!
A^\dagger_{ij}[\theta_{ij}] A_{ij}[\theta_{ij}] ) ,\nn
&& A^\dagger_{ij}[\theta_{ij}]  A_{ij}[\theta_{ij}]  = 2 [ 1  \!-\!
S_i^z S_j^z \nn
&& ~~~~~~~~~~~~~~~~~~ - ( e^{i\theta_{ij}} S_i^+ S_j^- +
e^{-i\theta_{ij}} S_i^- S_j^+ ) ] .\label{AKLT-in-1D}\ea
The phase rotation produces $ e^{i\theta_{ij}} S_i^+ S_j^- +
e^{-i\theta_{ij}} S_i^- S_j^+$ in Eq. (\ref{AKLT-in-1D}). Taking all
$\theta_{ij}=0$ gives back the usual AKLT Hamiltonian.


Having obtained the chiral extension of the one-dimensional AKLT
state, we consider some of its ground state properties and the
excitation energies using the single-mode approximation
(SMA)\cite{AAH}. For the ground state $|\{\theta_i \} \rangle$ in Eq.
(\ref{AKLT-wf}), the average of $S_i^- S_j^+$ is obtained from
$\langle\{\theta_i \} |S_i^- S_j^+ |\{\theta_i \}\rangle =
e^{i\theta_{ij}} \langle \mathrm{AKLT} | S_i^- S_j^+ |\mathrm{AKLT}
\rangle  = (1/3)e^{i\theta_{ij}} \langle \v S_i \cdot \v S_j
\rangle_0 $. Here the subscript $0$ refers to the average with
respect to the AKLT ground state. The chiral moment in $|\{ \theta_i
\} \rangle$ follows as $\kappa_{ij} = (1/3)\langle \v S_i \cdot \v
S_j \rangle_0 \sin \theta_{ij} = -(4/9)\sin\theta_{ij}$ for nearest
neighbours. The spin-spin correlation function is straightforward to
work out\cite{AAH}:

\ba  && \langle \{\theta_i \} |S_i^{x(y)} S_j^{x(y)} |\{\theta_i \}
\rangle = (1/3) \langle \v S_i \cdot \v S_j \rangle_0 \cos
\theta_{ij},\nn
&&\langle \{\theta_i \} |S_i^z S_j^z |\{\theta_i \} \rangle = (1/3)
\langle \v S_i \cdot \v S_j \rangle_0 , \nn
&& \langle \v S_i \cdot \v S_j \rangle_0 = 2\delta_{ij}+
4(1-\delta_{ij} ) (-1/3 )^{|i-j|}. \ea
The identity, $\langle \{\theta_i \} | S^x_i S^x_j |\{\theta_i \}
\rangle = \langle \{\theta_i \} | S^y_i S^y_j |\{\theta_i \}
\rangle$, is ensured by the global U(1) symmetry of the chiral
Hamiltonian, Eq. (\ref{AKLT-in-1D}). The exponential decay in the
spin-spin correlation persists for chiral AKLT states. The ensuing
SMA calculation, as well as the general argument for the invariance
of the energy spectra given earlier, confirms that the excitation gap
persists for non-zero chiral angles. The ground state is thus
non-magnetic, gapped, and possesses non-zero chiral moments.

\begin{figure}[t]
\begin{center}
\includegraphics[scale=0.6]{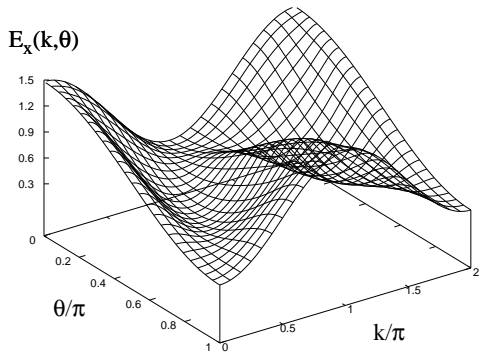}
\end{center}
\caption{SMA energies for $S^x_k$ excitations for the uniformly
chiral AKLT state $|\theta\rangle$ in 1D chain and $S=1$ as given in
Eq. (\ref{SMA-for-chiral-AKLT}). Plots are shown for  $0 \le k \le
2\pi$ and $0 \le \theta \le \pi$.} \label{1d-excitation}
\end{figure}

The structure factor for the uniform chiral AKLT state
$|\theta\rangle$, where $\theta_{ij} = \theta \times (i-j)$, can be
easily worked out. Denoting the structure factor in the AKLT state as
$s (k) =\langle S^z_{\kbar}S^z_k \rangle_0 = 2 (1-\cos k)/(5+3\cos
k)$, $\kbar\equiv -k$, we have

\be   \langle \theta |S_{\kbar}^{x(y)} S_k^{x(y)} |\theta \rangle =
[s (k\!+\!\theta ) + s (k\!-\!\theta)]/2 ,\ee
and $\langle \theta |S_{\kbar}^z S_k^z |\theta\rangle = s(k)$. The
average energy of the excited state $S^\alpha_k | \theta \rangle$
($\alpha=x,y,z$) is given by $\langle \theta | S^\alpha_i
H[\{\theta_i \}] S^\alpha_j |\theta\rangle$, which is equal to $\cos
\theta_{ij} \langle S^\alpha_i H S^\alpha_j \rangle_0$ for $\alpha =
x,y$, and $H$ is the AKLT Hamiltonian. For $S^z$, AKLT expressions
are obtained. The excitation energies in the SMA are given by

\be E_{x} (k, \theta ) = E_y (k,\theta) = {f(k\!+\!\theta) +
f(k\!-\!\theta) \over s(k\!+\!\theta) + s(k\!-\!\theta) }
,\label{SMA-for-chiral-AKLT}\ee
using $f(k) = (10/27)(1-\cos k)$. The excitation spectra are
displayed for $\theta$ ranging from $0$ to $\pi$ in Fig.
\ref{1d-excitation}. The SMA energies for $S^x_k |\theta\rangle $ and
$S^y_k |\theta\rangle$ possess symmetry under $k \leftrightarrow -k$,
while those for $S^+_k |\theta\rangle$ and $S^-_k |\theta\rangle$
will be given by $f(k\pm \theta)/s(k\pm \theta)$, respectively,
explicitly breaking the chiral symmetry. The SMA results are also in
accord with the general argument that all the energy eigenstates
remain in one-to-one correspondence through the rotation.

A string order parameter\cite{string-order} characterizes the
inherent antiferromagnetic spin-spin correlation in the AKLT state
better than the spin-spin correlation function itself, which has an
exponential fall-off with the separation. The string order parameter
in the chiral state is given by

\ba &&O^{x(y)}_{ij} = \notag \\ &&S^{x(y)}_i \exp\big[ i \pi \sum_{j<k<i}
\left(\cos\theta_k S^{x(y)}_k \!\pm\! \sin\theta_k S^{y(x)}_k \right)
\big] S^{x(y)}_j , \nn \ea
while the $z$-component of the string order is given by the usual
one: $O^z_{ij} = S^z_i \exp\big[ i \pi \sum_{j<k<i} S^z_k \big]
S^z_j$.
The averages of the string operators for $|\{\theta_i \} \rangle$ is
$\langle\{\theta_i \} | O^{x(y)}_{ij} |\{\theta_i \}\rangle = -(4 /9
)\cos \theta_{ij}$ and $\langle\{\theta_i \} | O^z_{ij} |\{\theta_i
\}\rangle = -4/9$. In particular for the uniform chiral phase, the
factor $\cos\theta_{ij} = \cos [\theta (i-j)]$ in the string order
reflects the extra pitch angle $\theta$ due to the helical spin
structure introduced by the DM interaction.
\\

\textbf{Higher-dimensional generalization}: Construction of chiral
AKLT ground states and the associated parent Hamiltonians are
possible in higher dimensions:

\begin{eqnarray}
&& |\chi \rangle = \prod_{\ij} \left(\Abar_{ij} [\theta_{ij}
]\right)^{M} |0\rangle \nn
&& H^\chi = \sum_{\ij} \overset{2S}{\underset{J=2S-M+1}{\sum}} K_J
\mathcal{P}^J_{ij}[ \theta_{ij} ], ~~K_J >0.
\label{HamiltonianChi}
\end{eqnarray}
Here $M=2S/z$ is determined by the value of the spin $S$ and the
lattice coordination number $z$. Each $\ij$ bond carries a bond angle
$\theta_{ij}$. With more bond variables than can be generated by the
set of site angles, the gauge rotation argument of the one dimension
does not readily apply in higher dimensions. An alternative proof is
given as follows.

The projector to the angular momentum-$J$ subspace\cite{AKLT,AAH}
$\mathcal{P}^J_{ij}$ in the AKLT Hamiltonian is constructed in terms
of the bond spin operator $\v J_{ij}^2 = (\v S_i + \v S_j )^2 = 2 S
(S+1) + 2 \v S_i \cdot \v S_j $. The replacement $S_i^+ S_j^- + S_i^-
S_j^+ \rightarrow e^{i\theta_{ij}} S_i^+ S_j^- + e^{-i\theta_{ij}}
S_i^- S_j^+$ in $\v J_{ij}^2$ produces $ \mathcal{P}^J_{ij}[
\theta_{ij} ] $ in Eq. (\ref{HamiltonianChi}).  To prove that $|\chi
\rangle$ is indeed the zero-energy ground state of $H^\chi$ in Eq.
(\ref{HamiltonianChi}), we will show that each projector
$\mathcal{P}^J_{ij}[ \theta_{ij} ]$ acting on $|\chi \rangle$
produces zero.

First write

\be \mathcal{P}^J_{ij}[ \theta_{ij} ] = V_j V_i \mathcal{P}^J_{ij}
V_i^\dagger V_j^\dagger , \label{ProjectorTheta}  \ee
where $V_i$ is the $U(1)$ rotation $V_i S_i^\pm V_i^\dagger = S_i^\pm
e^{\pm i \phi_i}$. We can choose $\phi_i$ and $\phi_j$ freely as long
as their difference is equal to $\theta_{ij}$.  Focusing on a given
bond $\ij$, the chiral ground state can be written out

\begin{widetext}
\ba && |\chi\rangle = \sum'_{m_i \!+\!n_i} \sum'_{m_j \!+\!n_j}
\cdots (a^\dagger_i)^{m_i}(b^\dagger_i)^{n_i} \left(e^{i\theta_{ij}}
a^\dagger_i b^\dagger_j - e^{-i\theta_{ij}} a^\dagger_j b^\dagger_i
\right)^M (a^\dagger_j)^{m_j}(b^\dagger_j)^{n_j} \cdots |0\rangle
\label{chi-expansion}
\end{eqnarray}
where the terms on the far left and far right stem from the product
of $\Abar_{pq}[\theta_{pq}]$ with only one end of $\langle pq
\rangle$ connected to either $i$ or $j$. The sum $m_i + n_i$ and $m_j
+ n_j$ are constrained to equal $2S-M$ in $\sum'$ above. Applying the
projector $\mathcal{P}^J_{ij}[ \theta_{ij} ]$ on $|\chi\rangle$ and
using relation \eqref{ProjectorTheta} we obtain

\begin{eqnarray}
&&\mathcal{P}^J_{ij}[ \theta_{ij} ]|\chi\rangle = \sum'_{m_i \!+\!
n_i} \sum'_{m_j\!+\!n_j}  \cdots  e^{i \psi(\phi_i,\phi_j)} V_j V_i
\mathcal{P}^J_{ij} \Bigl[ (a^\dagger_i)^{m_i }(b^\dagger_i)^{n_i
}\left(a^\dagger_i b^\dagger_j - a^\dagger_j b^\dagger_i \right)^M
(a^\dagger_j)^{m_j}(b^\dagger_j)^{n_j} \Big] \cdots |0\rangle . 
\label{ProjectorOnChiralState}
\end{eqnarray}
Here the phase $\psi(\phi_i,\phi_j) = (n_i - m_i) \phi_i+ (n_j -m_j)
\phi_j $ stems from the gauge transformation $V_i^\dagger V_j^\dagger
\cdots V_i V_j $ applied on the terms shown in Eq.
\eqref{chi-expansion}. The state shown inside the bracket $[ \cdots
]$ in Eq. (\ref{ProjectorOnChiralState}) have an expansion in terms
of states for which the total momentum $\v J_{ij}= \v S_i+ \v S_j$ on
the bond $\ij$ is less than or equal to $J_{max} = 2S - M$. Hence for
projectors $\mathcal{P}^J_{ij}[ \theta_{ij} ]$ with $J > 2S -M$, Eq.
\eqref{ProjectorOnChiralState} is zero. Since the whole argument
works for each bond $\ij$ we have proven that $H^\chi |\chi\rangle =
0$.

Although the proof holds for any bond angle configuration
$\{\theta_{ij}\}$ and for any dimension of the lattice, the gauge
transformation introduced in Eq. (\ref{ProjectorTheta}) is only a
local one, without the possibility to define the global unitary
operator constructed as the product of $V_i$'s. Hence, it is
generally not correct to associate $\sin \theta_{ij}$ with the local
average of the chirality except when we can decompose the bond angle
as the difference of the local angles, $\theta_{ij} = \theta_i -
\theta_j$.
\\
\end{widetext}

\textbf{Discussion}: In conclusion, we have identified a simple and
straightforward way to produce ground states of spins carrying
non-zero vector spin chirality. The key idea is to start with a spin
Hamiltonian whose ground state is non-chiral, and introduce
non-uniform phase twists of $S^+_i$ and $S^-_i$, but not of $S^z_i$.
The difference of the twist angle for nearby sites
$\theta_{ij}=\theta_i - \theta_j$ defines the degree of local vector
chirality. The Dzyaloshinskii-Moriya interaction also emerges in a
natural way, after implementing the non-uniform O(2) rotations on
the Hamiltonian without the DM interaction. A simple argument shows
that the ground state in the presence of the DM interaction will
generally possess non-zero vector chiral moments.

The well-known AKLT ground states of spins can be generalized in
this way, in both one and higher dimensions. The ground state
correlation properties for one-dimensional chiral AKLT states, in
particular, can be readily calculated as chiral rotations of the
known correlations for the non-chiral AKLT state. The excitation
energies for the uniformly chiral AKLT state is calculated within
the SMA and possess the gap which does not close as the chiral angle
is varied. We have in addition identified the string order parameter
appropriate for the linear chiral AKLT chain. Since the states we
constructed in this paper possess nonzero chiral moment, their
long-range ordering follows automatically. Construction of a
different kind of chiral state, without the chiral moment but only
long-range order in its correlations\cite{hikihara}, will be an
interesting challenge for the future.

For the experiments, insulating systems having the DM
interaction in addition to the Heisenberg superexchange,
such as the parent compound La$_2$CuO$_4$\cite{LSCO}, are
the likely places to find ground states with non-zero
vector spin chirality. While the lattice deformation
responsible for the presence of DM interaction can be
measured in the X-ray scattering, a direct, simultaneous
measurement of the spin chirality $\kappa_{ij}$ in the same
compound using the polarized neutron scattering\cite{onoda}
will highlight the correlation between the two phenomena
involving the lattice and the spin.

H. J. H. was supported by the Korea Research Foundation through Grant
No. KRF-2005-070-C00044. Insightful comments from Ki-Seok Kim are
gratefully acknowledged.

\end{document}